\documentclass[prd,amsmath,amssymb,floatfix]{revtex4}

\begin{document}

\title{What We Know and What we Don't Know About the Universe
\footnote{Keynote address at the International Workshop on
Astronomy and Relativistic Astrophysics, October 12-16 2003, Olinda, Brazil.}}

\author{Marcelo Gleiser}
\email{gleiser@dartmouth.edu}
\affiliation{Department of Physics and Astronomy
         Dartmouth College
				 Hanover, NH 03755, USA}

\date{\today}

\begin{abstract}
I present a non-technical and necessarily biased and incomplete
overview of our present 
understanding of the physical universe and its constituents, emphasizing 
what we have learned from the
explosive growth in cosmological and astrophysical data acquisition
and some of the key open questions that remain. The topics are
organized under the labels {\it space, time,} and {\it matter}.
Most bibliographical references are for the non-expert.
\end{abstract}

\maketitle

\section{Introduction}

It is true that the words ``cosmology'' and ``revolution'' have
appeared together in many reviews and lectures in recent years. 
Even though to a cynic it may seem that cosmologists produce
a lot of hot air, I claim that there is indeed reason for this 
excitement. Until the mid-sixties, cosmology was regarded with much
suspicion by many scientists, who thought it closer to metaphysics
than to physics. The reason for the skepticism was an appalling lack of
observational data, or at least consistent observational data. For
example, since the time of Hubble, the universe's age has varied from 
2 billion years (mcu below the known Earth's age even in 1929)
to over 20 billion. Its geometry has been bent and closed like the surface of a sphere,
flat as a tabletop, or bent and open like the surface of a saddle. (Of course,
generalized to 3 spatial dimensions.) Its material composition has also been
a mystery: which chemical elements are made in stars and which during 
the first moments after the ``bang''? What of other, exotic kinds of matter?
And what about the cosmological constant? 
Is it there or not? And if it is, what is it? \cite{hetherington}
Questions related to the shape, age, and material
composition of the cosmos have been at
the forefront of cosmology since Einstein's pioneering use of his theory
of relativity to the study of the universe as a whole.

Controversy is a good thing for science only when there are ways of resolving it.
This is why these are exciting times for cosmology. We have convincing
answers to many of the questions above, and are close to answering some
more. Of course, this does not mean that cosmology is approaching its
end. Quite the contrary, as with any mature area of science, 
new technologies and observational tools will continue to provide both answers 
and new questions and surprises.

\section{Space}

In astronomy and cosmology the yardstick is the light-year (ly), the distance
covered by light in 1 year, equal to $9.46\times 10^{12}$ km. That's
$63,200$ times the Earth-Sun distance, also known as astronomical unit (A.U.). Pluto
is at about $40$ A.U. from the Sun. Oort's cloud, the nursery of long-period
comets at the outskirts of the solar system, is at roughly $100,000$ A.U.,
about $1.58$ ly from the Sun. The nearest star, Proxima Centauri, is at $4.29$ ly
from the Sun. So, when people say that interstellar space is mostly empty you better believe it.
At least of ``visible'' stuff. It turns out, paraphrasing the 
fox in Saint Exup\'ery's {\it The Little Prince},
``what is essential [to the universe] is invisible to
the eye.'' Invisible but very real, as will be seen below.

Once we start thinking about other stars, it is inevitable to ask if there
are other planets as well. The thought that our
solar system is unique in some way brings nightmares to most (but not all) 
post-copernican scientists. Although the curiosity about extrasolar planets existed
for as long as modern astronomy (and before \cite{gleiser-pa}), the hunt started in earnest
in the mid-1990s when minute variations of stellar light frequency
could be detected due to improved observational technologies and computer data analysis software.
At the time of writing (January 2004), 104 planetary systems with 119 planets have been detected
[See www.obspm.fr/planets]. The curious lack of multiple planetary systems is most probably
due to the limitations of the observational methods; one should not consider this as any
indication of our uniqueness. NASA (Terrestrial Planet Finder Mission) and the European Space 
Agency (Darwin Mission) are formulating competing (possibly future collaborations) missions
with a projected capability of not only imaging Earth-size planets in the infrared 
but also of studying the chemical composition of their atmospheres. The goal is to explore the
possibility that these planets may harbor life. We may know the answer within a decade or two.
Or sooner, if programs like SETI are successful, or if we receive a visitor or message.
Improbable but not impossible.

Jumping farther out, the Milky Way -- our home galaxy -- is a spiral
containing about $300$ billion stars and
diameter of $100,000$ ly or $30$ kpc, where kpc stands for $10^3$ parsecs.
($1$ pc equals $ 3.3$ ly.) Strong evidence indicates that a giant black hole with a mass of
about $3$ million suns is at the dead center of the Milky Way, at Sagittarius A. Possibly most
other galaxies also host giant black holes at their centers, given the huge outpour of
radiation many produce at varying wavelengths. If you think $3$ million
solar masses is large think again; observations indicate that the huge galaxy M87 in the Virgo
Cluster host a black hole with $3$ {\it billion} solar masses.

Which brings me to clusters of galaxies and to the large-scale structure of the universe. 
The Andromeda galaxy, our nearest neighbor, is at about $2.5$ Mly (million light-years)
from the Milky Way. 
As we consider distances of millions of light-years, galaxies become the 
units by which we visualize the cosmos. When the Hubble Space Telescope or other large
terrestrial telescopes image deep space, the light they collect as point or small sources is coming
from huge galaxies millions and even billions of light-years away.
What we see as stars when we look at the night sky, large-scale
astronomers see as galaxies, each with millions or billions of stars. And what they
see is that galaxies may coexist in
groups or clusters due to their mutual gravitational attraction, like islands in an archipelago.
Furthermore, galaxies and cluster of galaxies are not randomly distributed across space;
they tend to collect in curved sheet-like surfaces, reminiscent of the shapes we see in bubble
baths. Only at distances of $100$ million Mpc or more these ``bubbly'' structures seem to
disappear, with the universe becoming smoother. 

One of the successes of modern cosmology was to shed partial light on the origin of such complex
large-scale structures. The answer involves of course gravity as the conductor, and three players:
ordinary, or baryonic, matter, the stuff made of protons as we are; a different kind of matter
known as dark matter \cite{bartusiak}, 
which we know exists from its gravitational pull on ordinary matter (there are
some dissenters, though);
and small energy inhomogeneities, the seeds that cause matter (first dark and then baryonic)
to start condensing. You may think of these energy inhomogeneities as lumps in an 
otherwise smooth cookie dough,
regions where the local density of energy (recall that matter and energy are treated equally in
relativity) varies from the average. So, we need three players. Computer simulations using them
at the right proportions (more about that later) reproduce quite well astronomical observations.  
But what is this dark matter
stuff, and what was the mechanism that generated the seeds that triggered structure and
galaxy formation? We will leave dark matter for later. It is believed that the inhomogeneities 
were originated during the earliest moments of cosmic existence, in a process called 
inflation\cite{guth}. During inflation, the universe expanded extremely fast. Even small 
fluctuations in its quite smooth energy were amplified enormously. Those are the culprits
for large-scale structure, some $100,000$ years after inflation ended. Although there is
no compelling model for inflation based on what we know of particles physics, the core idea
and its predictions seem to be in excellent agreement with observations. If nothing else,
the final answer will contain some of its elements. 

What of the shape of the whole cosmos? When Einstein wrote his paper on cosmology, he
assumed the universe was static and with the closed geometry of a 3-sphere. At the time,
there was no compelling reason to suppose otherwise, and spheres, since the time of the
Pythagoreans, have had an alluring effect on the minds of philosophers and scientists 
alike \cite{gleiser-du}.
A spherical universe is closed yet has no boundary, and all its points are equivalent.
(Imagine the surface of a perfect ball.) The question remained open until the advent of inflation
in the early-1980s. For when one talks of an expanding universe, one is referring to
an expanding geometry, as if the surface of a ball were inflating like a party balloon.
(But no one is blowing on it from the outside!) All points rush away from each other, and
distances increase at tremendous rates. Imagine then the balloon inflating, and focus
your attention on a patch on its surface. As it inflates, the patch becomes flatter.
An enormous amount of inflation leads to a practically flat patch. This patch, according
to inflationary cosmology, is where we live. Thus, inflation predicts a flat cosmos. This
prediction has been spectacularly confirmed by observations of the cosmic microwave
background (CMB), the fossil radiation from the time cosmic structure started to form \cite{turner}.

But our cosmos may be flat only locally. We can't say much about other folds of the cosmos, 
far removed from us.
In fact, we will never be able to know about those remote cosmic regions, as no signal
can be sent or arrive from there in able time: light is only so fast, and our 
cosmic patch has been around for about $13$ billion years. We cannot see beyond 
the flat island of $13$ ly radius we call our universe. Some models of inflation
even predict that the universe is really a ``multiverse,'' a bubbling soup of cosmoids
forever bursting in and out of existence \cite{linde}. We will have to wait a bit on this one.

\section{Time}

The big bang model is our current best description of cosmic history. It states that the
universe had a very hot and dense infancy, and that it has been expanding ever since.
There is now excellent concordance between different methods that estimate the age of
the universe. Here are a few: Measurements of the Hubble constant all the way to galaxies at
$500$ Mpc give $H_0= 72$ km/sec/Mpc in a universe with
$1/3$ matter and $2/3$ dark energy (more about this soon). 
This results in an age of $t_0 = 13\times 10^9$
years with an uncertainty of $1.5\times 10^9$ year \cite{freedman};
measurements of the CMB, independent of $H_0$, give $t_0=13.7\pm 0.2\times 10^9$ 
years\cite{wmap,knox}; Finally, the ages of the oldest stars in globular clusters are estimated to be
about $12.5\times 10^9$ years \cite{krauss}. A consistent age is $t_0=13.5\pm 1.0\times 10^9$
years. 

Within the big bang framework, the evolving cosmic history is a history of increasing
complexification: at the earliest times, matter was broken down to its smallest
components. To understand cosmic history is equivalent to recreate how the structures
that make up the material universe emerged. Complexification happened in stages, controlled
mostly by the cosmic temperature. At a given temperature, only certain particles or
their bound states (atomic nuclei, atoms, perhaps solitons) are in thermal equilibrium with
the ambient radiation. One may picture this as a love triangle, where photons
and possibly other light particles such as neutrinos (the radiation)
interact with matter particles which are trying to bind. If these interactions
proceed fast enough, the particles (or their bound states) are in equilibrium; otherwise, 
they are left behind as fossils of this era. Since different interactions have
different associated energy (temperature) scales, the cosmic history evolves in stages,
depending when a certain interaction goes out of equilibrium, from the most
to the least energetic. Thus, the early cosmic history went through at least
three stages: a particle era, a nuclear era, and an atomic era. The boundary
between the particle and nuclear eras is at about $10^{-5}$ sec, when quarks and gluons
combined to make the baryons, particles that interact by the strong
nuclear force such as protons, neutrons, and mesons \cite{gleiser-cp}. Since protons
have lifetimes of at least $10^{31}$ years, we can say that they are fossils from this
era. Once protons and neutrons were around,
they could start to bind into the lightest nuclei, H$_2$, H$_3$, He$_4$, He$_3$, and Li$_7$.
This happened when the universe was about 1 minute old. One of the
great triumphs of the big bang model is predicting successfully the abundances of these
nuclei. In particular, He$_4$ account for about $23$\% of the baryonic matter in the cosmos,
and hydrogen to about $75$\%.

The next step in the cosmic history is the formation of hydrogen atoms, at about $380,000$
years. This is when protons bind with electrons and photons are free to roam across space,
responding here and there to variations in the average energy density by having cold and
hot spots.
Measuring the properties of these photons, known as the CMB,
is equivalent to taking a snapshot of the universe at that early 
time \cite{smoot,mather}.
From then on, the universe became transparent to radiation. This means that it 
is impossible to try to study the cosmos before this time by direct measurements of
electromagnetic radiation of any wavelength; to probe the early universe we need to hunt for fossils.
Dozens of terrestrial missions and the satellites COBE and WMAP (and soon PLANCK) have produced
high precision measurements of the CMB, re-energizing modern cosmology and confirming
yet again the predictions of the big bang model. Measurements of the CMB allowed us to
confirm the flatness of space \cite{spergel}. (Recent claims to the contrary \cite{luminet},
remain a possibility, albeit a very improbable one.) They have also allowed us to learn when the
first stars were born (roughly at $200$ million years).

Understanding the nature of time remains one of the great challenges of physics. 
A classical theory of gravity (and thus the big bang model) predicts the existence of 
an ``initial singularity,'' the instant in time when the energy density 
reaches an infinite value and space collapses to a point. At the heart of the problem is a proper
formulation of a quantum theory of gravity, explaining its behavior
at very small distances and high energies. Although there are
models describing the origin of the universe (at least our local patch) by a quantum tunneling
event \cite{hawking,linde2}, the transition from a quantum to a classical universe
one described by the big bang model remains obscure. We need guidance from
particle physics to pick the proper theory to work with. String theory \cite{greene} and/or loop 
gravity \cite{smolin} have made much progress, but we still don't have a compelling connection
between their realm and that of 4-dimensional spacetime particle physics. 

\section{Matter}

It must be clear to the reader that within modern cosmology, treating
space, time, and matter as three separate topics is highly artificial. The three
are deeply intertwined. A point in case is the questions of the ``end of time'' \cite{gleiser-pa}.
Up to 1998, the standard big bang model had a very simple prediction:
if we know the total energy density of the universe 
($\rho_{\rm tot}$),
we can tell what will happen to it. The cosmic fate is controlled by the critical density, 
$\rho_{\rm crit} = 3H_0^2/8\pi G\simeq 10^{-29}$ g/cm$^3$, where $G$ is Newton's 
gravitational constant. It is convenient to define the ratio
$\Omega_i \equiv \rho_i/\rho_{\rm crit}$ for a given contribution to $\rho_{\rm tot}$.
Of course, $\rho_{\rm tot} = \sum_i \rho_i$,
and $\Omega_{\rm tot} = \rho_{\rm tot}/\rho_{\rm crit}$. If $\Omega_{\rm tot} > 1$,
the universe will recollapse in the distant future. Otherwise, it will continue its
expansion indefinitely. The discovery in 1998 that the universe is presently accelerating
blurred the clarity of this prediction \cite{livio}. The culprit of this acceleration
is called dark energy, a ghostly contribution to the total energy density (and pressure,
which makes 'dark energy' a somewhat imprecise name) spread out homogeneously
or nearly so across the universe.
Its net effect is to push the cosmic geometry apart, somewhat like an anti-gravitational
force. Measurements of distant supernovae and of the CMB give 
$\Omega_{\rm DE} \simeq 0.73\pm 0.04$. So, not only there is an unknown energy component
in the universe, but it is also the dominant one: CMB and other measurements place
$\Omega_{\rm tot} = 1.02\pm 0.02$, meaning that everything else must amount to
no more than $27$\% or so of cosmic stuff. Most probably this implies
that the universe is marginally flat and hence will expand forever. But what
if the universe were supercritical ($\Omega_{\rm tot} >1$? Without dark energy (pre 1998),
we would say it would recollapse in a finite time. But with dark energy things are more 
subtle \cite{krauss2,gleiser-pa}.

Dark energy is extremely difficult to pinpoint; its effects on the CMB are mostly relegated to very
large fluctuations. The two main contenders are the infamous cosmological
constant, created by Einstein to balance out his collapsing static spherical model,
and a hypothetical scalar field called quintessence \cite{caldwell}. The main difference
between a cosmological constant and a scalar field is in their pressures
and clustering properties. Scalar fields may have pressures of smaller absolute magnitude
and respond to variations in the gravitational field: {\it i.e.}, quintessence may cluster,
even if very subtly. The problem is that the models for quintessence are somewhat contrived,
at least from a particle physics viewpoint. If quintessence exists, it will almost certainly
not be a fundamental field, but a phenomenological order parameter or condensate of some
sort. The cosmological constant is not in much better shape. Quantum physics predicts that
all fields have a sort of fundamental inescapable jitter.
Since all that moves carry energy, this quantum jitter has an associated energy, called
zero-point energy \cite{weinberg,padmanabhan}. This energy is most certainly
related to the cosmological constant. Quintessence has nothing to say about this.
The problem is, if one would naively compute the
zero-point energy in the universe, the result would be a cosmological constant some $120$
orders of magnitude too large. In other words, we don't understand the cosmological
constant. Until we have a better grasp of it, dark energy will remain mysterious.

What then is the cosmic recipe, {\it circa} January 2004? Taking $\Omega_{\rm tot} =1$,
the total normal (baryonic) matter and dark matter components add up to $27$\%, or 
$\Omega_{\rm M} + \Omega_{\rm DM} = 0.27$. Nucleosynthesis constrains $\Omega_{\rm M} \simeq 0.04$,
leaving $\Omega_{\rm DM} \simeq 0.23$. That measurements of the distribution of dark matter in
galaxies and clusters of galaxies, as well as CMB measurements, concur very nearly
with this estimate speaks to the robustness of modern
observational and theoretical cosmology. It turns out that the stuff we and stars are made of
is by far the universes's subdominant energy component. Even this familiar component is only
partially accessible; some of ordinary matter is also ``dark'' in the sense that it does not
produce visible radiation. (But it produces other kinds of radiation, such as infrared, 
as we have seen in the hunt for extrasolar planets.) However, the bets are still off for
what is the dark matter component. Unless we are missing something very basic about 
gravity at large distances, dark matter is not of the ordinary kind. It may not be
of a single kind either: maybe several exotic particles and more complex objects
contribute to $\Omega_{\rm DM}$. A tremendous amount of effort is being dedicated to
disclose some of its nature, be it by direct detection (cryogenic detectors that would
``count'' the dark matter particles as they impact its collecting area or resonate in
special cavities) or by 
production in the laboratory (if dark matter is related to supersymmetry, it may
possibly be produced in particle accelerators very soon \cite{kane}.) Whatever dark matter
is, the stakes are high: we are talking of an unknown kind of matter, over six times more
abundant in the universe than ordinary matter. Both dark energy and dark matter promise
new physics ahead.

\section{CONCLUSIONS}

I hope this brief overview has given the reader at least a general
flavor of the excitement permeating present cosmological and astrophysical
research. We have learned a tremendous amount about the physical universe,
its material composition, its shape, and its age. Yet,
science teaches us that there will always be more to learn. In fact,
the old aphorism, ``the more we know the more we don't know'' seems
to apply very well here. As new tools and ideas open new vistas to the
physical universe, we stand in awe of nature's creativity and of our ability to
understand so much of it. To those that cynically comment on
the state of ``mystery'' in cosmology, ``too much dark stuff around,''
I say that this is precisely how science progresses: by teaching us
to accept ignorance as the main precondition for learning.
Only then we stand a chance of rationally answering some of the many questions we have,
and the many more that will surely come.
What we have learned so far speaks for itself.

\acknowledgements

I'd like to express my gratitude to the organizers of IWARA 2003, 
and in particular to Helio Coelho, Sandra Prado, and Cesar Vasconcellos for 
their invitation and wonderful hospitality. My research is sponsored
in part by a NSF grant PHY-0099543.

\end{document}